# Optical angular momentum derivation and evolution from vectorial field superposition

Liang Fang and Jian Wang

*Wuhan National Laboratory for Optoelectronics, School of Optical and Electronic Information, Huazhong University of Science and Technology, Wuhan 430074, Hubei, China*

Optical intrinsic angular momentum can be regarded as derivation from spatial dislocation of vectorial fields spinning or/and spiraling electric field vector. We employ fully vectorial formulation derivation to study all angular momentum contents of arbitrary superposed vectorial fields, including the longitudinal and transverse, spin and orbital (SAM and OAM) components. In nonparaxial orthogonal superposition fields, there inherently exists considerable spin-orbit shift from longitudinal SAM to OAM, and the whole spin flow manifests multiple-fold helical trajectories. Our presentation here provides an explicit insight into the derivation and evoation, the intrinsic correlations and the salient features of various angular momentum components.

Recently, extensive attention and academic interest have been increasingly attracted by photon's angular momentum, including spin and orbital angular momentum (SAM and OAM). The well-known longitudinal SAM and OAM with the same direction as the optical axis are associated with polarization states and spiral phase distribution of light beam, respectively, which is the salient intrinsic feature of light [1]. Photon's longitudinal OAM, as a new degree of freedom since pioneered by Allen and co-workers [2], has open a door to a large research fields with numerous applications in optical communication [3], quantum optics and information [4], optical tweezers and micromechanics [5], and even astronomy [6]. Apart from the longitudinal angular momentum, the classical transverse OAM as an extrinsic property of structured beam is associated with beam trajectory dependent of transverse coordinates of the beam centroid, similar to mechanical AM of classics particles [7]. Furthermore, the new-discovered transverse SAM has attracted a rapidly growing interest [8]. It arises as a result of the prominent longitudinal electric field, and exhibits unique features in sharp contrast to the usual longitudinal SAM, such as the effects of spin-momentum locking and lateral forces [9].

It is well known that vectorial fields are characterized by spatially inhomogeneous state of polarization, and manifest the full vectorial nature of electromagnetic wave [10]. They can be potentially applied to particle acceleration, microscopy, and sensing because of its unique properties [11]. Vectorial fields are supplied with higher-order solutions of vector Helmholtz equation [12]. In weakly guiding condition, fiber-guided modes carrying integer longitudinal SAM and OAM can be regarded as an attribution to a superposition with a phase shift of $\pi/2$ between two odd and even vector modes as eigenmodes of optical fiber waveguides [13]. However, when spatially arbitrarily superposing two vectorial fields, the resultant fields may exhibit distinct features on polarization state and spatial phase distribution that correspondingly spins and spirals the electric field vector. As a result, the longitudinal SAM and OAM as mean values are not confined to integers [14, 15]. Non-integer OAM also can be generated by using non-integer spiral phase plates [17], or with differential operators [18], astigmatic elements[19], and etc [20]. Especially, in high-contrast-index waveguides or in the case of nonparaxial propagation, there exists a nonnegligible angular momentum shift from SAM to OAM. It is analogous to the spin-to-vortex conversion of a paraxial beam in uniaxial anisotropic crystal or the polarization dependence of both SAM and OAM in nonparaxial case due to Berry-phase shift [16]. No-integer OAM and spin-orbit shift provide a discrete multi-dimensional state space for photons, which may find applications in quantum information processing, encryption and quantum digital spiral imaging [21]. Furthermore, in this case of nonparaxial fields, the transverse SAM increases sharply. The dominant ability of transverse spin is to achieve spin-controlled unidirectional propagation of light in nanofiber, surface plasmon-polaritons, and photonic-crystal waveguide [22], connected with the quantum spin Hall effect of light [23].

In this Letter, beyond paraxial approximation, we employ fully vectorial formulation derivation to comprehensively study the derivation and evolution of all optical angular mometum components from spatially arbitrarily superposing vectorial fields. We visualize the superposition to get arbitrary no-integer longitudinal SAM and OAM, and present the inherent spin-orbit shift and transverse SAM as a result of large longitudinal electric fields. We also show the whole spin flow as a combination of longitudinal and transverse SAM in the nonparaxial orthogonal superposition fields, and discuss the handedness of spin flow and the property of spin-momentum locking of transverse SAM.

Firstly, we derive the optical angular momentum components of general optical fields in fully vectorial formulation. Based on the canonical momentum expression [24], the longitudinal OAM density can be deduced in the cylindrically symmetric coordinate systems,

$$\mathbf{L}_z = -\frac{i\varepsilon_0}{2\omega}\left(\begin{array}{l}\psi_r^*\frac{\partial}{\partial\phi}\psi_r + \psi_\phi^*\frac{\partial}{\partial\phi}\psi_\phi + \psi_z^*\frac{\partial}{\partial\phi}\psi_z \\ -\psi_r^*\psi_\phi + \psi_\phi^*\psi_r\end{array}\right)\vec{\mathbf{e}}_z, \quad (1)$$

where $\omega$ is the angular frequency of light, $\varepsilon_0$ stands for permittivity in vaccum. In general, $\psi_r$, $\psi_\phi$, and $\psi_z$ are the radial, azimuthal, and longitudinal field components of vector fields, respectively. It represents the intrinsic vortex-dependent OAM that depends upon spatially varying phase distribution. As for the transverse OAM density, it can be given by

$$\mathbf{L}_\phi = -\frac{i\varepsilon_0 r}{2\omega}\left(\psi_r^*\frac{\partial}{\partial z}\psi_r + \psi_\phi^*\frac{\partial}{\partial z}\psi_\phi + \psi_z^*\frac{\partial}{\partial z}\psi_z\right)\vec{\mathbf{e}}_\phi. \quad (2)$$

which is origin-dependent and belongs to the extrinsic OAM. The SAM density with three components that is along the radial, azimuthal and longitudinal direction, respectively, can be written as

$$\mathbf{S} = -\frac{i\varepsilon_0}{2\omega}\left[\begin{array}{l}(\psi_\phi^*\psi_z - \psi_z^*\psi_\phi)\vec{\mathbf{e}}_r + (\psi_z^*\psi_r - \psi_r^*\psi_z)\vec{\mathbf{e}}_\phi \\ + (\psi_r^*\psi_\phi - \psi_\phi^*\psi_r)\vec{\mathbf{e}}_z\end{array}\right]. \quad (3)$$

The longitudinal SAM component $\mathbf{S}_z$ is well-known commonplace, of which the direction is determined by the polarization degrees of freedom. However, the unusual transverse SAM $\mathbf{S}_r$ and $\mathbf{S}_\phi$ are independent on the polarization of beam, but are determined by the longitudinal electirc field. Seeing Eqs. (1) and (3), the longitudinal OAM has an intrinsic longitudinal-SAM-dependence, associated with the origin of spin-orbital shift. Additionally, the time averaged energy density per unit length can be given as

$$W = \frac{1}{2}\varepsilon\left(|\psi_r|^2 + |\psi_\phi|^2 + |\psi_z|^2\right). \quad (4)$$

where $\varepsilon$ is permittivity of waveguide material.

When superposing two azimuthal-dependent vectorial fields with a phase shift of $\pi/2$ and a normalized energy allocation, one can get the arbitrary resultant vectorial field that propagates along the +z direction. It can be formulated by

$$\psi_{mn}(\varphi_1,\varphi_2,\gamma) = [\cos\gamma \cdot V_{mn}(\varphi_1) + i\sin\gamma \cdot V_{mn}(\varphi_2)]\exp[i(\omega t + kz)] \quad (5)$$

with

$$V_{mn}(\varphi_k) = E_r\cos(m\phi - \varphi_k)\cdot\vec{\mathbf{e}}_r + E_\phi\sin(m\phi - \varphi_k)\cdot\vec{\mathbf{e}}_\phi + iE_z\cos(m\phi - \varphi_k)\cdot\vec{\mathbf{e}}_z$$

$$(k = 1,2), \quad (6)$$

where $V_{mn}(\varphi)$ represents the general azimuthal-dependent vectorial field that is the solution of vector Helmholtz equation [12], $(r,\phi,z)$ describes the cylindrical coordinate, $m$ denotes the mode order or winding number, and $n$ the radial order, here we just consider the vectorial fields with first radial order, i.e. $n = 1$. $\varphi_1$ and $\varphi_2$ indicate the initial azimuthal orientation of two vectorial fields, respectively, which determines the spatial dislocation between two fields. $E_r$, $E_\phi$ and $E_z$ correspond to the radial, azimuthal and longitudinal electric field distribution of vectorial fields, respectively. The coefficients $\cos\gamma$ and $\sin\gamma$ $(0 \leq \gamma \leq \pi/2)$ allocate the normalized energy to two vectorial field components. Especially, when $m = 0$ and $\varphi_{1,2} = 0$, the vectorial field corresponds to the radially polarized mode $TM_{01}$, and when $m = 0$ and $\varphi_{1,2} = \pi/2$, it corresponds to the azimuthal polarized mode $TE_{01}$.

Insetting three vectorial electric field components of Eq. (5) into Eqs. (1) and (3), beyond paraxial approximation, we can express the longitudinal OAM density

$$L_z = \frac{\varepsilon_0\sin 2\gamma \cdot \sin\Delta\varphi}{4\omega}\left[m\left(E_r^2 + E_\phi^2 + E_z^2\right) + 2E_r E_\phi\right], \quad (7)$$

and the classical longitudinal SAM density

$$S_z = -\frac{\varepsilon_0}{2\omega}\sin 2\gamma \cdot \sin\Delta\varphi \cdot E_r E_\phi. \quad (8)$$

The transverse SAM density can be written by,

$$S_\phi = -\frac{\varepsilon_0}{\omega}E_r E_z\left[\cos^2\gamma\cos^2(m\phi - \varphi_1) + \sin^2\gamma\cos^2(m\phi - \varphi_2)\right] \quad (9)$$

$$S_r = \frac{\varepsilon_0}{2\omega}E_\phi E_z\left[\cos^2\gamma\sin 2(m\phi - \varphi_1) + \sin^2\gamma\sin 2(m\phi - \varphi_2)\right]. \quad (10)$$

Compared with longitudinal and transverse angular momentum density from Eqs. (7)-(10), the direction of longitudinal angular momentum is determined by spatial orientation dislocation $\Delta\varphi = \varphi_2 - \varphi_1$, i.e. the polarization handedness. However, the direction of transverse SAM does not dependent upon it. In the especial orthogonal superposition cases, i.e. $\Delta\varphi = \pm\pi/2$, there is no radial transverse SAM density, i.e. $S_r = 0$, while the azimuthal component becomes independent of azimuthal orientation, i.e. $S_\phi = -\varepsilon_0 E_r E_z/\omega$. The direction of transverse SAM only depends upon the propagagtion direction of light inherently regulating the phase correlation of radial and longitudinal electric field components. It gives rise to the phenomenon of spin-momentum locking belonging to the intrinsic property of transverse SAM, which makes sense that it enables spin-controlled unidirectional propagation of light [22].

The mean SAM and OAM in the unit of $1/\omega$ can be written as the ratio of the integral angular momentum to the averaged energy in the Minkowski expression form [25], as follows,

$$\frac{n_i^2\langle S_z\rangle}{\langle W\rangle} = \frac{n_i^2\iint S_z r dr d\phi}{\iint W r dr d\phi} = -\frac{\varsigma}{\omega}\sin 2\gamma \cdot \sin\Delta\varphi \cdot (1 - \xi), \quad (11)$$

for SAM, and

$$\frac{n_i^2 \langle L_z \rangle}{\langle W \rangle} = \frac{n_i^2 \iint L_z r dr d\phi}{\iint W r dr d\phi} = \frac{1}{\omega}\sin 2\gamma \cdot \sin \Delta\varphi \cdot \left[m + \varsigma(1-\xi)\right], \quad (12)$$

for OAM, where the $n_i$ is the refractive index of optical medium, subscript $i$ corresponds to different waveguide layer, and $\xi$ represents the amount of spin-orbit shift in orthogonal superposition state, and given by

$$\xi = 1 - \varsigma \frac{2\int E_r E_\phi r dr}{\int \left(|E_r|^2 + |E_\phi|^2 + |E_z|^2\right) r dr}. \quad (13)$$

For HE$_{ml}$ modes, $\varsigma = -1$, and the signs of $E_r$ and $E_\phi$ are opposite; whereas for EH$_{ml}$ modes (including TM$_{01}$ and TE$_{01}$ modes with $m=0$), $\varsigma = 1$, and $E_r$ and $E_\phi$ have the same signs. In weakly guiding approximation or paraxial propagation, generally, $E_z \simeq 0$, $E_r \simeq E_\phi$, and thus $\xi \simeq 0$, when $\Delta\varphi = \pm\pi/2$ and $\gamma = \pi/4$, the topological charges of OAM and SAM is $\ell = \pm(m-1)$, and $s = \pm 1$ for orthogonal superposition of HE$_{ml}$ modes, whereas $\ell = \mp(m+1)$, and $s = \pm 1$ for orthogonal superposition of EH$_{ml}$ modes $(m \neq 0)$; and especially, $\ell = \mp 1$, and $s = \pm 1$ for TM$_{01} \pm i$TE$_{01}$ modes $(m=0)$.

From expressions above, the values of mean OAM can be taken within a continuous range, and the mean SAM can take arbitrary values from $-1$ to $+1$. The angular momentum has an intrinsic spin-orbit tangle for non-planar optical wave. The spin-orbit shift becomes remarkable in the nonparaxial case, because of the non-negligible longitudinal electric field $E_z$. The amount of spin-orbit shift can be quantified by Eq. (13). The ratio of the sum of integral longitudinal SAM and OAM to the averaged energy can be given by

$$\frac{n_i^2(\langle S_z \rangle + \langle L_z \rangle)}{\langle W \rangle} = \frac{1}{\omega} m \sin 2\gamma \cdot \sin \Delta\varphi. \quad (14)$$

It is just dependent on the superposition states and the winding order $m$. In any superposition states, it is conserved for the total longitudinal angular momentum.

Analogously, we give the expressions in terms of the ratio of the integral transverse OAM and SAM to the averaged energy, as follows,

$$\frac{n_i^2 \langle L_\phi \rangle}{\langle W \rangle} = \frac{rk}{\omega}, \quad (15)$$

for transverse OAM,

$$\frac{n_i^2 \langle S_\phi \rangle}{\langle W \rangle} = -\frac{\pi}{\omega} \frac{\int E_r E_z r dr}{\int \left(|E_r|^2 + |E_\phi|^2 + |E_z|^2\right) r dr}, \quad (16)$$

for azimuthal SAM,

$$\frac{n_i^2 \langle S_r \rangle}{\langle W \rangle} = 0, \quad (17)$$

for radial SAM that has no comtribution to the integral transverse SAM.

To visually illustrate the resultant vectorial fields in arbitrary superposition states on the basis of vector beams, the superposition state $\psi_{mn}(\Delta\varphi,\gamma)$ can be described with an unit semisphere. Here the azimuth-dependent angle $\varphi_2$ is fixed as $\varphi_2 = \pi/2$, ($V_{mn}(\pi/2)$ corresponds to the odd vectorial field), and $0 \leq \Delta\varphi < 2\pi$, $0 \leq \gamma < \pi/2$. For instance, we plot the polarization states and longitudinal OAM values in arbitrary superposition states on the basis of HE$_{21}$ mode, as shown in Fig. 1. The spin-orbit shift in this case is $\xi = 0.12$. Three projections of the superposition states in Cartesian coordinate system can be derived by

$$\psi_{21}^1 = \cos\gamma \cdot \cos\Delta\varphi \cdot \psi_{21} \quad (18)$$

$$\psi_{21}^2 = \cos\gamma \cdot \sin\Delta\varphi \cdot \psi_{21} \quad (19)$$

$$\psi_{21}^3 = \sin\gamma \cdot \psi_{21} \quad (20)$$

and

$$\left|\psi_{21}^1\right|^2 + \left|\psi_{21}^2\right|^2 + \left|\psi_{21}^3\right|^2 = \left|\psi_{21}\right|^2 \quad (21)$$

where $\psi_{21}^1$, $\psi_{21}^2$, $\psi_{21}^3$ indicate the vector mode components HE$^o$, HE$^e$, and $i$HE$^o$, respectively.

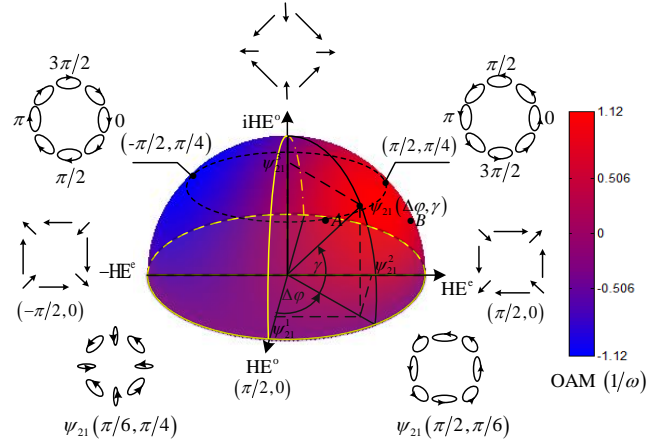

Fig. 1 Arbitrary polarization states and OAM evolution from arbitrary superposition states $\psi_{21}(\Delta\varphi,\gamma)$ on the basis of HE21 mode with the amount of spin-orbit shift $\xi = 0.12$. Position A and B correspond to superposition states $\psi_{21}(\pi/6,\pi/2)$ and $\psi_{21}(\pi/2,\pi/6)$, respectively.

On the semisphere, the yellow circle on the bottom denotes uniform vector field $V_{21}(\varphi_1)$ with an unit amplitude. The yellow semicircle across vertical axis indicates the superposed vectorial field of $\cos\gamma \cdot HE^o + i\sin\gamma \cdot HE^o$. These superposition states do not carry SAM and OAM. Beyond them, any other states on the semi-spherical surface would spin and spiral the spacial electric field vector. It gives rise to spatial-variant asymmetry of polarization ellipticity, analogous to the combination of hybrid states of polarization in Refs. (26). It thereby induces intriguing arbitrary non-integer values of mean SAM and OAM. The colour map on semi-spherical surface represents different mean OAM values. The yellow semicircle divide the semi-spherical surface into two components. The polarization orientation of the first quadrant is characterized by left handedness, and that of the second quadrant is right handedness. In the center of two quadrants, i.e. $\psi_{21}(\pi/2,\pi/4)$ and $\psi_{21}(-\pi/2,\pi/4)$ as orthogonal superposition states, the polarization states display uniform ellipse distribution with an azimuthal symmetry. This nonparaxial superposed vectorial field carrying non-integer OAM can be found in high-contrast-index waveguides, for instance, hollow ring-core fiber [27], where exists considerable spin-orbit shift. It degrades into the fully circular polarization carrying integer SAM and OAM in the weakly guiding fibers due to the nearly identical radial and azimuthal field components. In reverse, when $\gamma = 0$ or $\pi/2$, and $\eta = \pm 1$, the resultant vectorial fields can describe fiber-guided vector modes and can be combined by two purely circularly polarized OAM modes [28]. Note that when $\varphi_2$ varies, it changes the azimuthal angle of the overall superposition fields, but does not affect the investigation on SAM and OAM. It is worth pointing out that there is a noticeable difference between the resultant fields expressed by Eq. (5) and the fields described in higher-order Poincaré sphere [28]. In higher-order Poincaré sphere, two OAM modes with opposite SAM and OAM states serve as the mode bases. Nevertheless, there is only one azimuthal orientation dependence linked to two mode bases. However, the resultant vectorial fields here are on the basis of two vector modes, both of them are azimuthal orientation dependent. It can be equivalent to presentation of higher-order Poincaré sphere if two azimuthal angles $\varphi_1$ and $\varphi_2$ here are constrained with $\Delta\varphi = \pi/2$ in Eq. (5). The case of superposition based on two uniform vectorial fields in this Letter has an additional degree of freedom in terms of optical polarization states.

Finally, we investigate the whole SAM of the orthogonal superposition fields $\psi_{mn}(\pm\pi/2,\pi/4)$ in the nonparaxial case. For nonparaxial fields, $TM_{01}$ and $TE_{01}$ modes are split, whereas even and odd HE/EH modes are degenerated associated with the effective refractive index. Considering the combination of longitudinal and transverse SAM density, we plot the spin flow trajectories of uniform vector field $TM_{01}$ that is closed loop as shown in Fig. 2(a). Note that the $TE_{01}$ vectorial field does not have transverse SAM, because of no longitudinal electric field, as shown in Fig. 2(b). We further present the spin flow of orthogonal superposed fields $\psi_{mn}(\pm\pi/2,\pi/4)$ on the basis of $HE_{11}$ and $HE_{21}$ modes, respectively, as shown in Figs. 2(c)-2(f), in contrast, we also give the spacial phase distribution of these fields that reflect optical OAM states. These spin trajectories as a combination of longitudinal and transverse SAM are characterized by helixes. It arises as a result of phase offset of transverse SAM lobes along the propagation around the optical axis based on the defination of SAM from Eqs. (3) and (9). The fold number of spin trajectories equals to the azimuthal ordor $m$ of vectorial fields. The spin trajectories of orthogonal superposed fields with $\psi_{mn}(+\pi/2,\pi/4)$, $(m \neq 0)$ are characterized by right-handed helix with the same direction as the propagation direction aligned with the linear momentum k, as shown in Figs. 2(c) and 2(e). However, for superposed fields with $\psi_{mn}(-\pi/2,\pi/4)$, $(m \neq 0)$, the spin trajectories manifest left-handed helix with the reverse direction as the propagation direction, disaligned with the linear momentum k, as shown in Figs. 2(d) and 2(f). Significantly, in all cases above, the direction of transverse SAM remains the same, and is locked to the linear momentum k, which is the intrinsic property of transverse SAM [9].

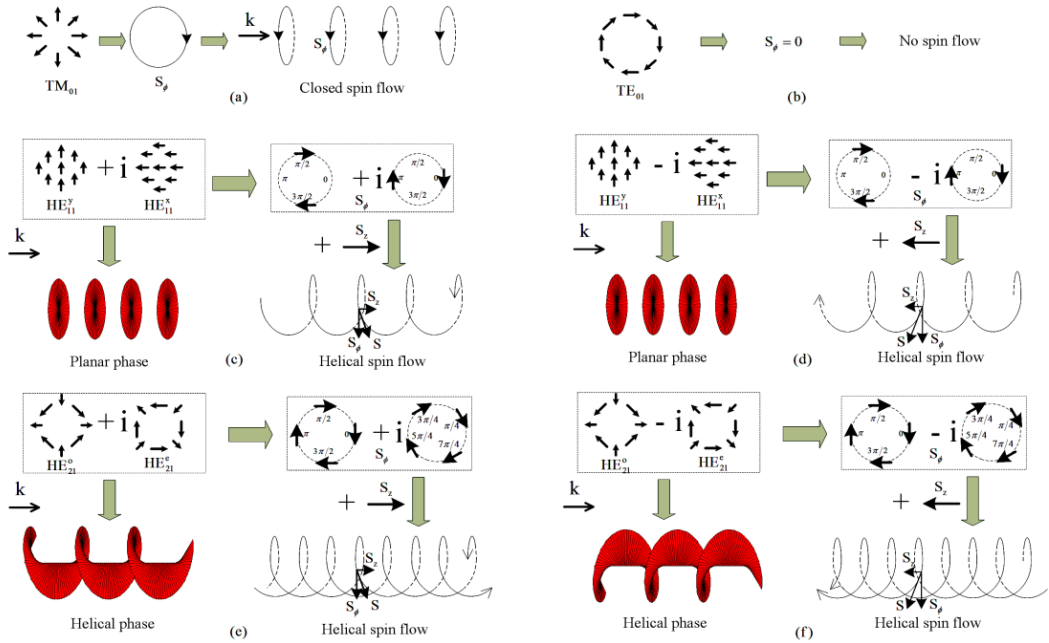

Fig. 2 Spin flow trajectories of (a) $TM_{01}$ mode and (b) $TE_{01}$ mode. Spacial phase distribution and spin flow trajectories of superposed fields for (c) $\psi_{11}(+\pi/2,\pi/4)$, (d) $\psi_{11}(-\pi/2,\pi/4)$, (e) $\psi_{21}(+\pi/2,\pi/4)$, and (f) $\psi_{21}(-\pi/2,\pi/4)$.

In conclusion, the resultant field by arbitrarily superposing vectorial fields has been comprehensively investigated in terms of various optical anguler momentum in this Letter. For a nonparaxial superposed field, there are remarkably inherent spin-orbit shift and transverse SAM due to the large longitudinal electric field. We can obtain arbitrary non-integer OAM and fractional SAM by arbitrarily superposing the vectorial fields. The whole spin flow of $TM_{01}$ and higher-order superposing fields manifests muple-fold helical trajectories, of which the fold number is the same as the azimuthal order. This helical optical spin can enrich optical force and torque as new degree of freedom that enables to exploit the next generation of photonic traps. Our revelation and presentation in this Letter provide a systematic physical insight into optial angular momentum derived from vectorial field superposition, including the longitudinal and transverse, SAM and OAM components, spin-orbit shift and evolution, and the property of transverse SAM. It may facilitate the development of optical vectorial fields and optical angular momentum in fundamental studies and various applications, such as optical communications, optical manipulation, and quantum application, etc.

This work was supported by National Basic Research Program of China (973 Program, 2014CB340004); National Natural Science Foundation of China (NSFC) (11274131); Program for New Century Excellent Talents in University (NCET-11-0182).